\documentclass[12pt]{article}

\AtBeginDocument{
  \addtocontents{toc}{\normalsize}
}

\pdfoutput=1

\usepackage{amsmath,amssymb,amsfonts,amsthm,amscd,mathrsfs}
\usepackage{xcolor}
\definecolor{darkblue}{rgb}{0.1,0.1,.7}
\usepackage[colorlinks, linkcolor=darkblue, citecolor=darkblue, urlcolor=darkblue, linktocpage]{hyperref} 
\usepackage[square, comma, compress,numbers]{natbib}
\usepackage[]{version}
\usepackage[]{graphicx}
\usepackage[]{latexsym}
\usepackage{geometry}
\usepackage[all,cmtip]{xy}
\usepackage[margin=10pt,font=small,labelfont=bf]{caption}
\usepackage{ifthen}
\usepackage{tikz}
\usepackage{array,setspace,mathrsfs,amsfonts,yfonts,dsfont,bbm,colonequals}
\usepackage{dsfont} 
\usepackage{xspace}
\usepackage{subcaption}

\usepackage{accents}
\newlength{\dhatheight}

\newcommand{\reef}[1]{(\ref{#1})}

\def\eps{\epsilon}
\newcommand{\beq}{\begin{equation}} 
\newcommand{\eeq}{\end{equation}}
 
\def\nn{\nonumber} 
 
\def\bR {\mathbb{R}} 
 
\def\bN {\mathbb{N}} 
 
\def\calG {{\cal G}} 
 
\def\calF {{\cal F}}

\def\half{{\textstyle\frac 12}}

\def\ge{\geqslant}
\def\le{\leqslant}

\def\nn{\nonumber}

\def\eps{\epsilon}

\newcommand{\be}{\begin{equation}}
\newcommand{\ee}{\end{equation}}
\def\ba{\begin{array}}
\def\ea{\end{array}}

\newcommand{\D}{\Delta}
\newcommand{\Df}{{\Delta_\phi}}

\numberwithin{equation}{section}
\begin{document}

\vspace*{-.6in} \thispagestyle{empty}
\begin{flushright}
CERN PH-TH/2017-098
\end{flushright}
\vspace{1cm} {\large
\begin{center}
{\bf Cut-touching linear functionals in the conformal bootstrap}
\end{center}}
\vspace{1cm}
\begin{center}
{\bf Jiaxin Qiao$^{a,b}$, Slava Rychkov$^{a,b}$}\\[2cm] 
{
\small
$^a$ Laboratoire de physique th\'eorique,\\ D\'epartement de physique de l'ENS,
\'Ecole normale sup\'erieure, PSL Research University,\\ 
Sorbonne Universit\'es, UPMC Univ.~Paris 06, CNRS, 75005 Paris, France\\
$^b$  CERN, Theoretical Physics Department, 1211 Geneva 23, Switzerland
\normalsize
}
\end{center}

\vspace{4mm}
\begin{abstract}
The modern conformal bootstrap program often employs the method of linear functionals to derive the numerical or analytical bounds on the CFT data. These functionals must have a crucial ``swapping" property, allowing to swap infinite summation with the action of the functional in the conformal bootstrap sum rule. Swapping is easy to justify for the popular functionals involving finite sums of derivatives. However, it is far from obvious for ``cut-touching" functionals, involving integration over regions where conformal block decomposition does not converge uniformly. Functionals of this type were recently considered by Maz\'a\v{c} in his work on analytic derivation of optimal bootstrap bounds. We derive general swapping criteria for the cut-touching functionals, and check in a few explicit examples that Maz\'a\v{c}'s functionals pass our criteria.

\end{abstract}

\vspace{1cm}
\hspace{0.3cm} \noindent May 2017

\newpage

{
\setlength{\parskip}{0.05in}
\renewcommand{\baselinestretch}{0.7}\normalsize
\tableofcontents
\renewcommand{\baselinestretch}{1.0}\normalsize
}


\setlength{\parskip}{0.1in}

\section{Introduction}

Any conformal field theory (CFT) is characterized by the spectrum of local primary operators and by their operator product expansion (OPE) coefficients, called collectively the CFT data. This data has to satisfy a consistency conditions following from the OPE associativity. The program of constraining or solving the CFT data using the OPE associativity is known as the conformal bootstrap 
\cite{Ferrara:1973yt,Polyakov:1974gs,Mack:1975jr,Belavin:1984vu}. The conformal bootstrap equations are mathematically well defined and can be studied on a computer \cite{Rattazzi:2008pe}. This approach has led in the last 10 years to a wealth of rigorous numerical results about CFTs, many of which are currently out of reach of analytical methods.

Following \cite{Rattazzi:2008pe}, the numerical conformal bootstrap analysis is usually formulated in terms of \emph{linear functionals}, as we will now review (see also \cite{Rychkov:2016iqz,Simmons-Duffin:2016gjk}). Let us specialize to the case of a one-dimensional (1d) CFT, by which we mean here a theory of local operators in 1d whose correlation functions transform covariantly under the fractional linear transformations $x\to (ax+b)/(cx+d)$. These form the group ${\rm SL}(2,\bR)$, the 1d counterpart of the group of global conformal transformations in $d$ dimensions. 
Consider in such a theory a four point (4pt) correlation function of a primary operator $\phi$ of scaling dimension $\Delta_\phi$. Conformal invariance constrains this correlator to have the form:
\beq
\langle \phi(x_4)\phi(x_1)\phi(x_2)\phi(x_3)\rangle = |x_{12}|^{-2\Delta_\phi}|x_{34}|^{-2\Df} \calG(z)\,,
\eeq
where $x_{ij}=x_i-x_j$ and $z={x_{12} x_{34}}/({x_{13}x_{24}})$ is the conformally invariant cross ratio. We are assuming the operators to be cyclically ordered on the conformally compactified real axis (as appropriate in 1d). If we put the operators at $0,z,1,\infty$ then we have\footnote{Defining as usual $\phi(\infty)=\lim_{z\to\infty} |z|^{2\Df}\phi(z)$.}
\beq
\langle \phi(\infty)\phi(0)\phi(z)\phi(1)\rangle = z^{-2\Df} \calG(z)\,.
\eeq
The function $\calG(z)$ is initially defined on the real interval $0<z<1$ (its analytic continuation will be discussed below). Since we are dealing with a 4pt function of four identical operators, this function satisfies on this interval the following \emph{crossing relation}
\beq
z^{-2\Df}\calG(z) = (1-z)^{-2\Df}\calG(1-z)\,,
\label{crossing}
\eeq
which is the simplest example of a conformal bootstrap equation. Near the endpoints of the interval, we have asymptotic behavior
\beq
\calG(z)\to 1\quad (z\to 0),\qquad \calG(z)\sim \frac{1}{(1-z)^{2\Df}}\quad (z\to 1)\,.
\label{4ptz}
\eeq
This is given by the unit operator contribution in the OPE $\phi\times\phi$, and is clearly consistent with the crossing relation.

Furthermore, the function $\calG(z)$ can be expanded into conformal blocks \cite{DO1,DO2,DO3}:\footnote{For $d>1$ conformal blocks depend also on the spin of the exchanged primary and on the second cross ratio $\bar z$.}
\beq
\calG(z)=\sum_{i=0}^\infty p_i\, G_{\D_i}(z),\qquad G_\D(z)=z^\D {}_2 F_1(\D,\D,2\D;z)\,.
\label{CBexp}
\eeq
Here $\Delta_i$ are the scaling dimension of all primary operators appearing in the OPE $\phi\times\phi$. The $p_i$ are the squares of the OPE coefficients. We have $\Delta_0=0$, $p_0=1$ corresponding to the unit operator. We will assume that our theory is unitary. In such theories all subsequent operators satisfy the unitarity bound $\Delta_i>0$. Also the OPE coefficients are real in unitary theories, implying $p_i\ge 0$. We will also assume for simplicity that the spectrum of operators is discrete without accumulation points, so that there is a finite number of operators below any fixed dimension. This assumption is not crucial and can be relaxed; see appendix \ref{sec:cont}.

Eqs.~\reef{crossing}, \reef{CBexp} can be rewritten as a sum rule
\beq
\sum p_i F_{\D_i}(z)=0,\qquad F_\D(z)=z^{-2\Df}{G_\D(z)}{} - {(1-z)^{-2\Df} }{G_\D(1-z)}\,.
\label{cross}
\eeq
($F$'s also depend on $\Df$ but we omit this dependence in the notation.)
This equation can be used to put constraints on the allowed unitary CFT spectra \cite{Rattazzi:2008pe}. The strategy is to look for a linear functional
\beq
\omega:f\mapsto \omega(f)
\eeq
which satisfies the conditions
\beq
\omega(F_{\D_0})>0\,,\qquad  \omega(F_{\D_i})\ge 0\qquad(i>1)\,. 
\label{positve}
\eeq
for all $\Delta_i$ in some putative spectrum. Applying $\omega$ to \reef{cross} and \emph{assuming we can swap the order of summation with the action of the functional} (a nontrivial requirement since the number of operators in the OPE is always infinite), we have
\beq
\sum p_i\, \omega(F_{\D_i})=0\,,
\eeq
which is impossible in view of \reef{positve} and of $p_0=1$, $p_i\ge0$ in unitary CFTs. The putative spectrum is thus ruled out.

The linear functionals used in \cite{Rattazzi:2008pe} and in essentially all subsequent numerical work were linear combinations of a finite number of derivatives at the midpoint $z=\half$:
\beq
\omega(f)=\sum_{n\le N} c_n f^{(n)}(\half)\,.
\label{funNum}
\eeq  
As we will review below, for these functionals the above-mentioned crucial swapping assumption is very easy to justify, basically because the conformal block decomposition converges uniformly near $z=\half$.\footnote{This is also true for functionals in \cite{Echeverri:2016ztu} using points away from $z=\half$ but within the region of uniform convergence.}

Recently Maz\'a\v{c} \cite{Mazac:2016qev} introduced a new class of linear functionals. Unlike the functionals of \cite{Rattazzi:2008pe}, his functionals involve integrals of $f$ over regions of the $z$ space approaching the analyticity cuts where the conformal block decomposition seizes to converge. The use of such ``cut-touching" functionals raises anew the problem of justifying swapping. His functionals are very interesting because, as explained in his paper, they lead to an analytic understanding of some optimal conformal bootstrap bounds previously conjectured by extrapolating numerical results. He verified the relevant conditions analogous to \reef{positve} in his paper. However, he has not discussed nor even mentioned swapping. This is an unfortunate gap in his otherwise beautiful analysis. 

 Although some amends were made in the online presentation \cite{Mazac-seminar},
we consider this issue not fully clarified, and sufficiently important to dedicate this short note to it. The functionals like in \cite{Mazac:2016qev}, or even more complicated ones, may well become widespread in the conformal bootstrap.
Anticipating these developments, we will show here a minimal standard of rigor in dealing with such functionals. Following this standard is important to ensure that the results are technicalIy correct. While in this note we focus on the case $d=1$, the standard we impose has a natural extension to $d>1$, and we hope it will be followed there as well.

We start in section \ref{sec:anal} by discussing the analytic continuation of the 4pt function and of its conformal block decomposition into the plane of complex $z$. In section \ref{sec:func} we formalize the swapping property (along with the more obvious finiteness) which the linear functionals used in the conformal bootstrap must have. For the usual functionals these properties are trivially verified. Then in section \ref{sec:mazac} we turn to the cut-touching functionals. We derive some general criteria guaranteeing that such functionals obey finiteness and swapping. In section \ref{sec:relation} we use our criteria to prove swapping for the functionals used in \cite{Mazac:2016qev}, at least for the particular cases of low-lying $\Df$ where \cite{Mazac:2016qev} provides sufficient details. The general case remains incomplete. 

Appendix \ref{sec:toy} contains a simple counterexample, showing that taking swapping for granted and proceeding formally can lead to manifestly wrong results. Appendix \ref{sec:cont} deals with the case when the operator spectrum is continuous or has accumulation points.

\section{Analytic continuation}
\label{sec:anal}

The function $\calG(z)$, while originally defined on the interval $0<z<1$, allows an analytic continuation into the complex plane of $z$ with cuts along {$(-\infty,0)$ and} $(1,+\infty)$ (``cut plane"). Analytic continuation is provided by the series \reef{CBexp}. Clearly, the individual terms in the series are analytic functions in the cut plane. In addition, the series converges in the cut plane.
To show this latter fact, it is convenient to work in the $\rho$ coordinate \cite{Pappadopulo:2012jk,Hogervorst:2013sma}:
\beq
\rho(z)=\frac{z}{(1+\sqrt{1-z})^2},\quad z(\rho)=\frac{4\rho}{(1+\rho)^2}
\eeq
The cut $z$ plane is thus mapped to the disk $|\rho|<1$. 

Consider the series \reef{CBexp} transformed to the $\rho$ coordinate:
\beq
\tilde \calG(\rho)=\sum p_i\, \tilde G_{\D_i}(\rho)\,,\qquad
\tilde \calG(\rho) = \calG(z(\rho)),\qquad \tilde G_\D(\rho) = G_\D(z(\rho))\,.
\label{Grho}
\eeq
Using hypergeometric function identities, we have \cite{Hogervorst:2013sma}:
\beq
\tilde G_\D(\rho) = (4\rho)^\Delta {}_2F_1(\half,\Delta,\Delta+\half;\rho^2)\,.
\label{HogG}
\eeq
We will not actually need this exact formula but three properties of conformal blocks that it implies:
\begin{gather}
\tilde G_\D(r)\ge 0 \qquad(0\le r<1)\,,\\
\tilde G_\D(r) = O\Bigl(\log\frac 1{1-r}\Bigr)\qquad (r\to 1)\,,
\end{gather}
and
\beq
|\tilde G_\D(r e^{i\theta})|\le \tilde G_\D(r)\,.
\label{est}
\eeq
The first property is obvious, the second is a standard hypergeometric asymptotics.
The last property follows by expanding the hypergeometric function in \reef{HogG} in a power series and noticing that all coefficients are positive (if $\D\ge0$ as demanded by unitarity).

Now we can finish the convergence proof. For real $\rho=r$, $0<r<1$, the function $\tilde \calG(\rho)$ is finite, and all terms in its series \reef{Grho} are positive, so the series does converge. On the other hand, for complex $\rho=r e^{i\theta}$, $r<1$, each term in the series is dominated by its value at $\rho=r$. So the series converges in the disk $|\rho|<1$. The original series \reef{CBexp} then converges in the cut $z$ plane. 

\begin{figure}
\centering
\includegraphics[width=0.5\textwidth]{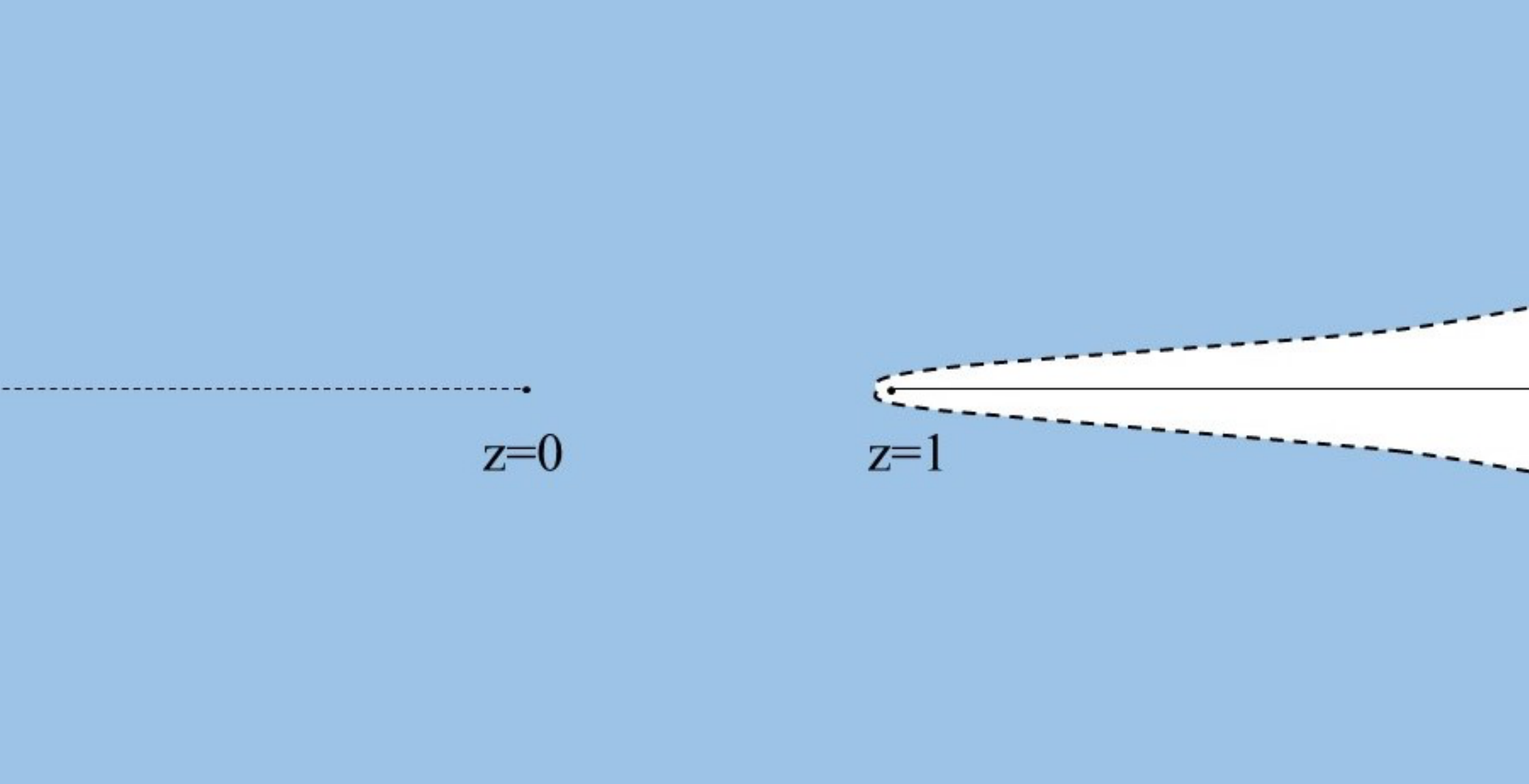}
\caption{A region where the series \reef{CBexp} converges uniformly (the image of the disk $|\rho|\le 1-\eps$ in the $z$ plane).}
\label{fig:Geps}
\end{figure}

The above argument is a 1d adaptation of the general $d$-dimensional argument from \cite{Pappadopulo:2012jk}.
The argument is robust and can be extended in several directions. For example, the same argument shows that the convergence in any subdisk $|\rho|\le 1-\eps$ is uniform. It is also easy to argue that the convergence in any such subdisk is exponentially fast \cite{Pappadopulo:2012jk}. For a precise formulation, let $\kappa$ be any real number in the range $1-\eps<\kappa<1$. Then there is a constant $C$ such that for any $\Delta_*$ the tail of the series \reef{Grho} corresponding to summing over $\Delta_i\ge \Delta_*$ satisfies the uniform bound:
\beq
\Bigl|\sum_{\Delta_i\ge \Delta_*} p_i\, \tilde G_{\D_i}(\rho)\Bigr |\le C \kappa^{\Delta_*}\,\qquad\text{for all}\ |\rho|\le 1-\eps\,.
\eeq
The r.h.s. of this bound becomes exponentially small for large $\Delta_*$.\footnote{One can also put $\kappa=1-\eps$ at the cost of making the constant $C$ grow as a power of $\Delta_*$ \cite{Pappadopulo:2012jk}, but we will not need this sharper estimate here.}
The subdisks $|\rho|\le 1-\eps$ are mapped onto the subregions of the $z$ plane shown in figure \ref{fig:Geps}. In any such subregion the series \reef{CBexp} converges uniformly and exponentially fast.

{ Also notice that the cut through $(-\infty,0)$ is present in the above argument only because the factors $z^\Delta$ and $\rho^\Delta$ in the conformal blocks have this cut. The convergence is not spoiled by the presence of this cut. In fact the argument proves that the function $\calG(z)$ can be analytically extended through this cut, and one can circle around the origin through a second, third etc sheet. The same is of course true for the cut $(1,+\infty)$ because the function $\calG(z)$ is crossing symmetric, Eq.~\reef{crossing} (or because we can equivalently run the argument around $z=1$). In this way one can explore the full domain of analyticity of $\calG(z)$, which is an infinitely sheeted Riemann surface if $\Delta_\phi$ is an irrational number. In this work we will stay on the first sheet.}

A comment is in order concerning the origin of the positivity property of the conformal blocks and of their power series coefficients, which played an important role in the above proof. In terms of the 4pt unction, passing to the $\rho$ coordinate corresponds to mapping it conformally to the configuration 
\beq
\label{eq:4ptrho}
\langle \phi(-1)\phi(-\rho)\phi(\rho)\phi(1)\rangle\,.
\eeq
For real $\rho<1$, the new configuration is reflection positive. This explains why all terms in the power series expansion of $\tilde \calG(\rho)$ have to be positive \cite{Pappadopulo:2012jk}. 

\section{Functionals: general considerations}
\label{sec:func}

\begin{figure}
\centering
\includegraphics[width=0.5\textwidth]{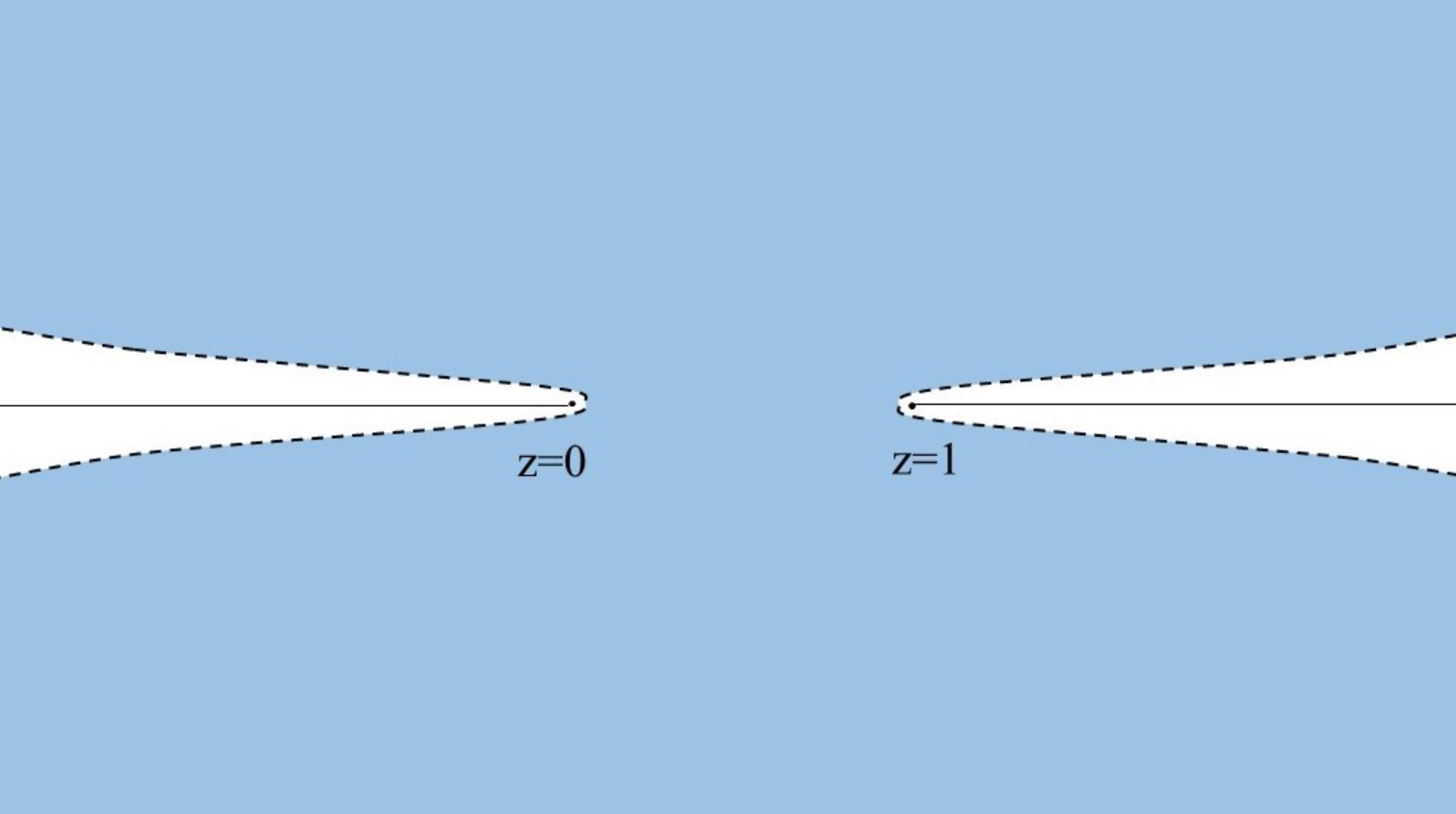}
\caption{A region of uniform convergence of the series in the crossing relation \reef{cross}.}
\label{fig:Feps}
\end{figure}

So let us go back to the crossing relation \reef{cross} satisfied by a 4pt function of some 1d CFT. Based on the discussion of the previous section, the following facts are true:
\begin{itemize}
\item 
Functions $F_\Delta$ are analytic in the cut plane
\item 
The series converges in the cut plane\,.
\item 
The convergence is uniform in the subregions where both conditions $|\rho(z)|<1-\eps$ and $|\rho(1-z)|<1-\eps$ are satisfied (see figure \ref{fig:Feps}).
\end{itemize}

We would like to consider linear functionals $\omega(f)$ which have the following two properties:
\begin{enumerate}
\item[P1.] (Finiteness) $\omega(F_\Delta)$ is finite for any $\Delta\ge0$\,.
\item[P2.] (Swapping) For any possible 4pt function of an operator of dimension $\Df$, Eq.~\reef{cross} implies that
\beq
\sum p_i\, \omega(F_{\D_i})=0\,,
\label{toshow}
\eeq
the series converging in the usual sense.
\end{enumerate}

It's important to emphasize that the functional should be defined not just on the functions $F_\D$ and on their finite linear combinations, but on a wider class of functions.\footnote{In this respect the notation of Eqs.~(2.8) and (2.22) of \cite{Mazac:2016qev} is confusing, while that in \cite{Mazac-seminar} is OK.} This class should at the very least include the functions $\calF_{ \Delta_*}(z)$ which will be introduced shortly.

In practice, the functional $\omega(f)$ will be given by some sort of integral or a combination of derivatives and property P1 should be relatively easy to check, especially given that the conformal blocks in 1d are known explicitly. 
Property P2 is more tricky. It can be ``derived" by applying functional $\omega(f)$ to both sides of \reef{cross}. However this is formal since it requires interchanging the action of the functional with infinite summation. Sometimes this formal argument is easy to justify, sometimes more work is needed. We will see examples in a second.

Assuming that P1 holds, the strategy to establish P2 is as follows. Split \reef{cross} into two parts (we switch from summing over $i$ to summing over the discrete set of occurring $\Delta$'s):
\beq
\sum_{\Delta< \D_*} p_\Delta F_{\D}(z)+\calF_{ \Delta_*}(z)=0,\quad \calF_{\Delta_*}(z)\equiv \sum_{\Delta\ge \D_*} p_\Delta F_{\D}(z)\,.
\label{calF}
\eeq
Now we can apply $\omega$ and get:
\beq
\label{interchange}
\sum_{\Delta<\D_*} p_\D \omega(F_\D)+\omega(\calF_{\Delta_*})=0\,.
\eeq
Notice that here we interchanged the functional with a \emph{finite} summation, which is always a legal operation.
Furthermore, the function $\calF_{\Delta_*}$ goes to zero in the cut plane as $\D_*\to\infty$, uniformly so in the regions shown in figure \ref{fig:Feps}. So we may expect that, under wide conditions on the functional $\omega$, 
\beq
\omega(\calF_{\Delta_*})\to 0\qquad(\D_*\to\infty)\,.
\label{Fsmall}
\eeq
If we can show this rigorously, then \reef{toshow} follows and we are done. This is what it takes to justify the formal argument.

Let us consider two examples where \reef{Fsmall} is immediate.

\begin{figure}[h]
\centering
\includegraphics[width=0.4\textwidth]{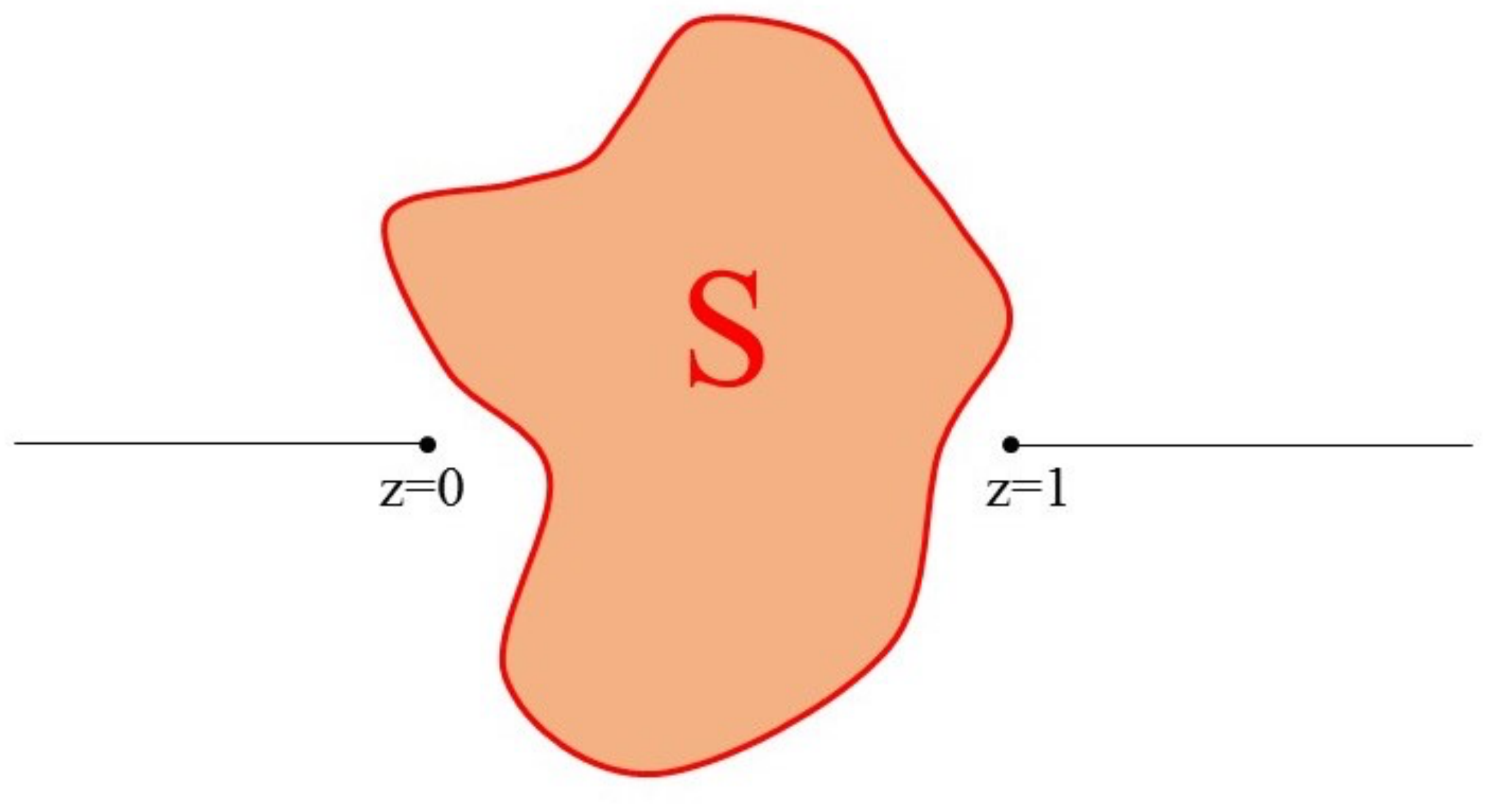}
\caption{Support of integration in the functional of Example 1.}
\label{fig:ex1}
\end{figure}

{\bf Example 1.} Suppose the functional $\omega$ is given by an integral over some integrable measure $d\mu$ whose support $S$ is a bounded set, which is fully contained in the cut plane and does not touch the cuts (see figure \ref{fig:ex1}):
\beq
\omega(f)=\int_S d\mu\, f(z)\,.
\label{ex1}
\eeq
Then \reef{Fsmall} follows trivially from the uniform convergence of \reef{cross} on $S$.

{\bf Example 2.} Suppose the functional $\omega(f)$ is a derivative of a finite order $n$ at a point $z_0$ lying strictly inside the cut plane:
\beq
\label{ex2}
\omega(f)=f^{(n)}(z_0)\,.
\eeq
This example can be reduced to the previous one, by representing the derivative via Cauchy's formula as a contour integral over a circle fully contained in the cut plane. 

Clearly, a finite linear combination of derivatives will do as well. The functionals \reef{funNum} used in the numerical bootstrap belong to this class. The simplicity of verification of \reef{toshow} in this case explains why it was left implicit in the literature. For example, the authors of Ref.~\cite{Pappadopulo:2012jk} carefully established the convergence of the conformal block decomposition in the cut plane and stated that this puts the numerical conformal bootstrap results on ``mathematically solid ground". What they had in mind was a kind of the above argument.

\section{Cut-touching functionals}
\label{sec:mazac}
\begin{figure}[h]
\centering
\includegraphics[width=0.8\textwidth]{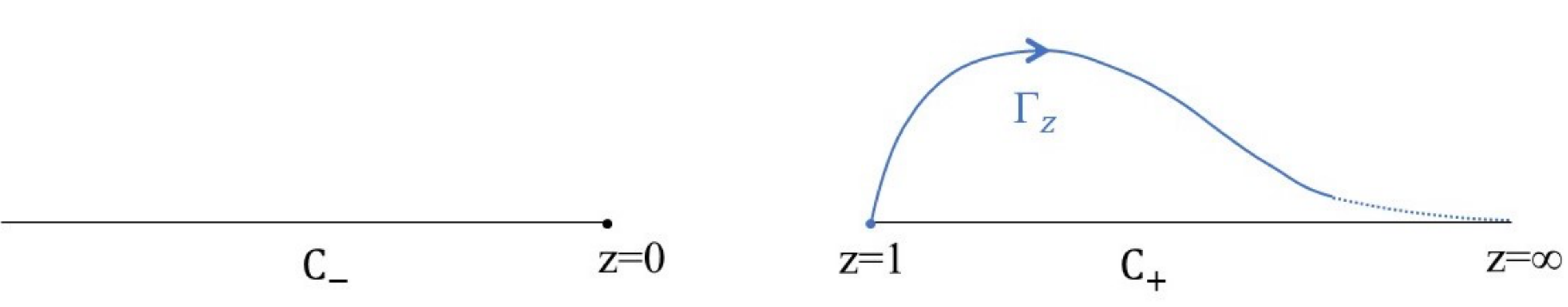}
\caption{Contour $\Gamma_z$\,. Also shown are the two cuts of the cut plane.}
\label{Gammaz}
\end{figure}

We will now consider a functional of the following form:
\beq
\omega(f)={\rm Im}\int_{\Gamma_z} dz\,H(z) f(z) \,,
\eeq
where $H(z)$ is a fixed analytic function in the upper half-plane. The function $f(z)$ on which the functional acts is also assumed analytic in the upper half-plane (in fact it will be analytic in the cut plane). The contour $\Gamma_z$ starts at $z=1$ and ends at $z=+\infty$, as shown in figure \ref{Gammaz}. 
Of course since the functions are analytic we may deform the contour. For example, we may want to make it run along the cut. Such contour deformations may be useful in actual explicit calculations, but for the proof of properties P1, P2 it will be convenient to keep the contour in the bulk of the upper half-plane, touching its boundary only at two points as shown.

\begin{figure}[h]
\centering
\includegraphics[width=0.8\textwidth]{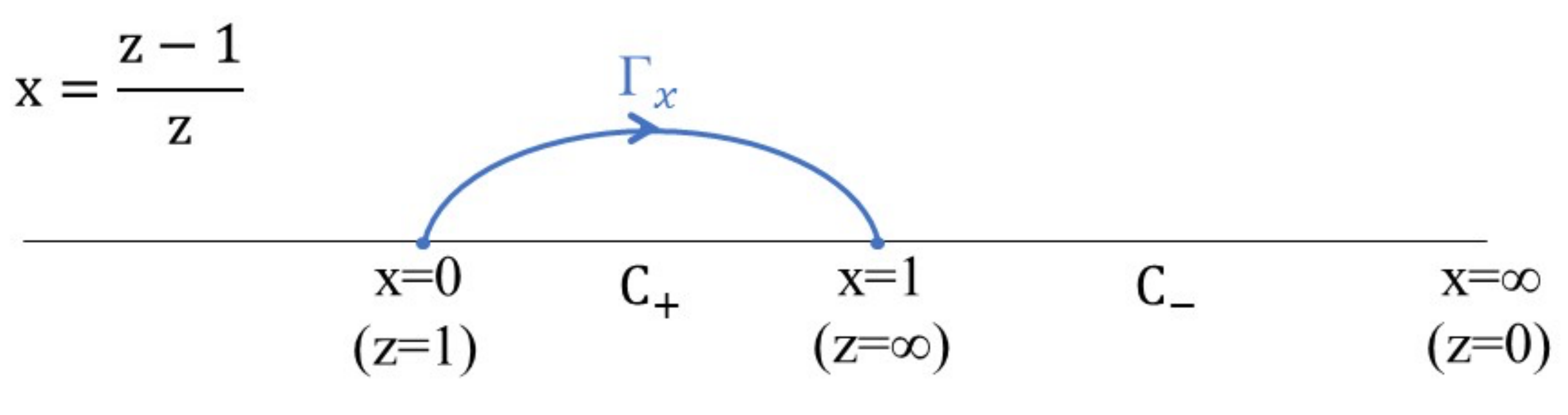}
\caption{Contour $\Gamma_x$\,. Also shown are the images of the two cuts of the cut plane under the transformation from $z$ to $x$. The function $h(x)$ and the functions $f(z(x))$ on which the functional is evaluated will be analytic in the upper half-plane.}
\label{Gammax}
\end{figure}

As in \cite{Mazac:2016qev}, let us pass from the coordinate $z$ to coordinate 
\beq
x(z)=\frac{z-1}{z},\quad z(x)=\frac 1{1-x}\,.
\eeq
The upper half-plane of $z$ is mapped to the upper half-plane of $x$ with points $0,1,\infty$ and the contour mapped as in figure \ref{Gammax}. It is equivalent but more convenient to analyze the functional in terms of the $x$ coordinate:
\beq
\omega(f)={\rm Im}\int_{\Gamma_x}dx\, h(x) (1-x)^{-2\Df} f(z(x))\,.
\label{omh}
\eeq
The function $h(x)=H(z(x)) z'(x) (1-x)^{2\Df}$ may have some singularities on the real axis but we will assume it is analytic in the upper half-plane. The factor $(1-x)^{-2\Df}$ is factored out for future convenience, as in \cite{Mazac:2016qev}. We will assume in our analysis that contour $\Gamma_x$ approaches $x=0,1$ not tangentially to the real axis.

It's clear that for such functionals the proof of swapping given above for Examples 1,2 cannot be applied. The problem is that the convergence of the series \reef{cross} near $z=1,+\infty$ (which map to $x=0,1$) is not uniform. To establish \reef{Fsmall}, we will need to understand how $\calF_{\Delta_*}$ behaves near these points. The condition for swapping, whatever it is, will depend in a nontrivial way on $\Df$ and on the asymptotics of $h(x)$ near $x=0,1$. Our goal here will be to work out this condition.

\subsection{Finiteness}
\label{sec:fin}

To check finiteness, we need to estimate how $F_\Delta(z(x))$ behaves near $x=0,1$. For $x\to0$ we have $z\approx 1+x$,
\beq
G_\D(z)=O (\log1/|x|),\qquad G_\D(1-z)=O(|x|^{\Delta})\,.
\eeq
For $x=1+\eps$, $\eps\to0$ we have $z\approx -1/\eps$. To estimate $G_\D(z)$ we pass to the $\rho$ coordinate:
\beq
\rho = \rho(z) \approx \rho(-1/\eps)\approx -1+2\sqrt{\eps}\,.
\eeq
For the crossed channel we have:
\beq
\rho' = \rho(1-z) \approx \rho(1/\eps)\approx -1-2i \sqrt{\eps}\,.
\eeq
Since we are assuming that $\eps$ is not parallel to the real axis we have
both
\beq
|\rho|,|\rho'|=1-O(\sqrt{|\eps|})\,. 
\eeq
Using the estimate \reef{est}, we have
\beq
|\tilde G_\D(\rho)|\le \tilde G_\D(|\rho|) = O(\log 1/|\eps|)\,,
\eeq
and analogously for $\tilde G_\D(\rho')$. 

Combining the above estimates for $G_\D$'s we can estimate $F_\D$. We have:
\beq
x\to 0:\quad(1-x)^{-2\Df} F_\D(z(x))=O (\log1/|x|) + O(|x|^{\Delta-2\Df})=O(|x|^{-2\Df})\,,
 \eeq
where we used $\D\ge0$, $\Df>0$. Further
\begin{align}
x=1+\eps:\quad (1-x)^{-2\Df} F_\D(z(x))&= z^{2\Df} F_\D(z(x)) \nn\\
&= O(G_\D(z)) + O(G_\D(1-z)) = O(\log 1/|\eps|)\,.
 \end{align}

Finiteness will hold if the following integrals involving these bounds are absolutely convergent:
\beq
\int dx\, h(x) |x|^{-2\Df}
\label{L1-1}
\eeq
should be integrable near $x=0$, while
\beq
\int dx\, h(x) \log \frac{1}{|1-x|}
\label{L1-2}
\eeq
should be integrable near $x=1$. Both integrals are assumed taken along the contour $\Gamma_x$.

\subsection{Swapping}

Let us split the contour $\Gamma_x$ into three parts, two ``end parts", one close to $x=0$ and one close to $x=1$, and the ``bulk part". As $\D_*\to\infty$, the function $\calF_{\D_*}(z(x))$ goes to zero uniformly (and exponentially fast) on the bulk part. So that part of the integral can be made arbitrarily small by choosing a sufficiently large $\Delta_*$.

On the end parts, we will estimate $\calF_{\D_*}$ as follows. First of all we bound all terms by absolute value:
\beq
|\calF_{\D_*}(z(x))|\le {|z|^{-2\Df}}{| \calG_{\D_*}(z)|} + {|1-z|^{-2\Df}}{|\calG_{\D_*}(1-z)|}\,.
\eeq
Here $\calG_{\D_*}$ is the tail of the conformal block decomposition, defined as $\calF_{\D_*}$ in \reef{calF} but summing over $G_\Delta$.

Using \reef{est}, we can estimate these tails by the whole function $\calG$ evaluated at the absolute value of the $\rho$ coordinate:
\beq
| \calG_{\D_*}(z)|\le \tilde\calG(|\rho(z)|)\le \frac {const.}{(1-|\rho(z)|)^{4\Df}}\,.
\label{estend}
\eeq
The second estimate can be understood for example by estimating the 4pt function using the OPEs $\phi(\rho)\times \phi(1)$ and $\phi(-\rho)\times \phi(-1)$ in \reef{eq:4ptrho}. Alternatively it just follows from the second of the asymptotics \reef{4ptz}. There is also an analogous estimate for $\calG_{\D_*}(1-z)$ with $\rho(1-z)$. 

One might think that the estimate \reef{estend} is too crude. However, we will need this estimate only near the endpoints of the contour, and there it's basically best possible.

Now, for $x\to0$, $z\approx 1+x$ (recall that the contour is not along the real axis) this strategy gives us:
\beq
|\rho(z)|\approx 1-const.\sqrt{|x|}
\eeq
and
\beq
\tilde \calG(|\rho(z)|)=O(|x|^{-2\Df}),\qquad \tilde\calG(|\rho(1-z)|)=O(1)\,.
\eeq
On the other hand, for $x=1+\eps$, using the estimates on $\rho$'s from the previous section we find:
\beq
\tilde \calG(|\rho(z)|), \tilde \calG(|\rho(1-z)|) =O(|\eps|^{-2\Df})
\eeq

We now combine these estimates on $\tilde \calG$ to get estimates on $\calF_{\D_*}$. We have:
\begin{align}
x\to 0:&\quad(1-x)^{-2\Df} \calF_{\D_*}(z(x))=O (|x|^{-2\Df})\,, \nn\\
x=1+\eps:&\quad (1-x)^{-2\Df} \calF_{\D_*}(z(x))= z^{2\Df} \calF_{\D_*}(z(x))\nn \\
&\qquad= O(\calG_\D(z)) + O(\calG_\D(1-z)) = O(|\eps|^{-2\Df})\,.
\label{estFD*}
 \end{align}
It's crucial for what follows that the r.h.s. of these estimates does not depend on $\Delta_*$. 

Suppose now that the following integrals of $h(x)$ against these bounds are absolutely convergent:
\begin{gather}
\int dx\, h(x) |x|^{-2\Df}\quad\text{over the part of $\Gamma_x$ near $x=0$}\,,\label{L2-1}\\
\int dx\, h(x) |1-x|^{-2\Df} \quad\text{over the part of $\Gamma_x$ near $x=1$}\,.\label{L2-2}
\end{gather}
Then we claim that the swapping property holds.

To show \reef{Fsmall} we argue as follows. Pick any $\delta>0$. Take the end parts of the contour sufficiently short so that those parts of the integral, for any $\Delta_*$, are smaller in absolute value than $\delta$. This is possible by the conditions \reef{L2-1},\reef{L2-2}.
The bulk part of \reef{Fsmall} tends to zero as $\Delta_*\to\infty$, since the integrand uniformly converges to zero there. We conclude that the large $\Delta_*$ limit of \reef{Fsmall}, in absolute value, is smaller than $\delta$. Since $\delta$ is arbitrary, the limit is zero. This completes the proof.

Notice that while \reef{L2-1} is identical to \reef{L1-1}, the other condition is stricter than \reef{L1-2}. In other words, the fact that the functional is finite on each $F_\Delta$ does not yet guarantee swapping. 

We would like to finish this section with the following comment. The problem of justifying the swap of integration and summation is of course standard in mathematics. One powerful result is Lebesgue's dominated convergence theorem. There are several reasons why we chose not to appeal to it in our exposition, but to deduce everything from scratch. First, Lebesgue's theorem is very general (it deals with an almost everywhere convergent sequence of measurable functions), and it's not good practice to shoot sparrows with a cannon. Second, if we did appeal to this theorem, we could eliminate but the paragraph following the conditions \reef{L2-1},\reef{L2-2}. The estimates \reef{estFD*} would still have to be derived (``dominated convergence"), and this is what constitutes anyway the bulk of our argument. Finally, we believe that there is an added value in seeing what actually goes into the proof. 

\section{Relation to the work of Maz\'a\v{c}}
\label{sec:relation}

The cut-touching functionals from the previous section are closely related to the functionals constructed in \cite{Mazac:2016qev}, with the purpose to give an analytic proof of a certain optimal bootstrap bound involving operators of dimension $\Df\in \bN/2$. Let us review this connection in detail. 

\begin{figure}[h]
\centering
\includegraphics[width=0.6\textwidth]{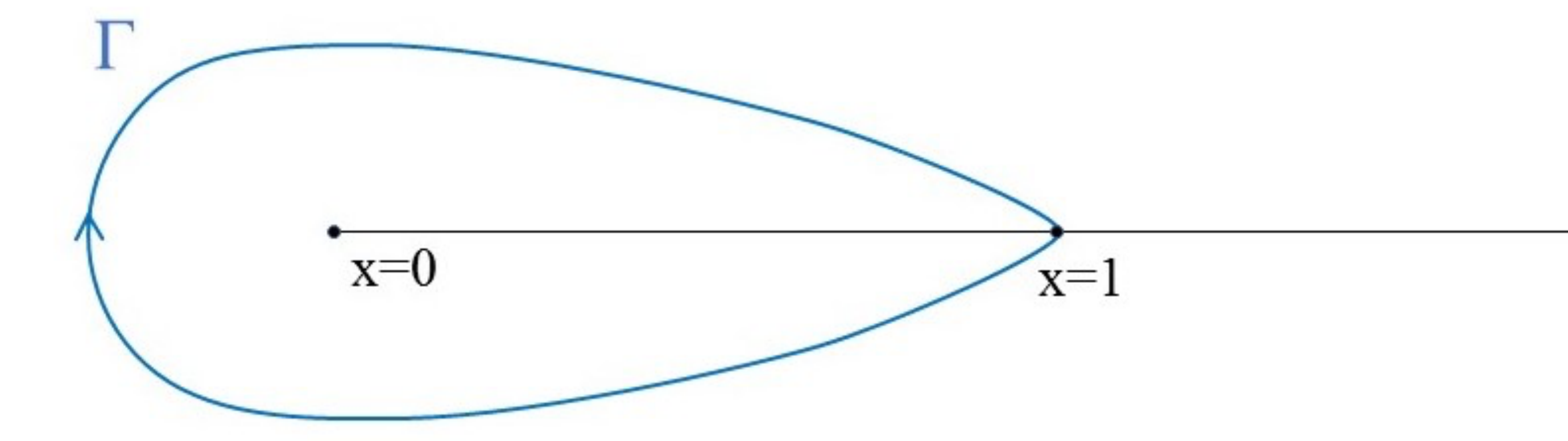}
\caption{The contour used in the definition of basis functionals \reef{basis}.}
\label{Mazac-contour}
\end{figure}

Maz\'a\v{c} begins by considering a family of basis functionals of the form
\beq
\omega(f)=\frac 1{2\pi i} \int_{\Gamma}dx\, h(x) (1-x)^{-2\Df} f(z(x))\,,
\label{basis}
\eeq
with $h(x)=p_n(x)$ a Legendre polynomial. The function $f$ is assumed analytic in the cut plane. He chooses the contour $\Gamma$ to run as in figure \ref{Mazac-contour}, staying away from the point $x=0$. Conditions for the finiteness\footnote{Ref.~\cite{Mazac:2016qev} actually works out $\omega(F_\Delta)$ for all functionals in closed form. So their finiteness is not in doubt.
We will still discuss finiteness for completeness, but our focus is on justifying swapping.} and swapping of these functionals can be examined exactly as above. It's clear that only conditions at $x=1$ need to be imposed. The finiteness condition 
\reef{L1-2} is satisfied. On the other hand, the swapping condition \reef{L2-2} is not satisfied, because $p_n(1)\ne 0$.

That's not a problem because he does not act with the basis functionals themselves on the sum rule. 
Instead he considers their linear combinations, corresponding to 
\beq
h(x)=\sum_n a_n p_n(x)\,.
\label{seriesh}
\eeq
The coefficients $a_n$ have to be chosen so that several conditions are satisfied. First of all, since his goal is to prove an optimal bootstrap bound, the functional has to be \emph{extremal}, which means that it has to satisfy certain positivity conditions closely related to \reef{positve}. These conditions have been discussed in detail in his work and we will not discuss them here.

Then the functional has to have the swapping property. This was not actually discussed in \cite{Mazac:2016qev}. Near $x=1$ we must have
\reef{L2-2}, which should arise as a result of cancelation between various terms in the sum defining $h(x)$.\footnote{The talk \cite{Mazac-seminar} (29m30s) cites the condition $h(x)=O((x-1)^\Df)$ as needed ``for the functional to be defined on infinite sums of blocks bounded at infinity". This is not far from our condition \reef{L2-2}, although a bit stronger than necessary. We emphasize however that the functional has to be not just ``defined", but has to satisfy Eq.~\reef{Fsmall} from which the swapping property follows.}

\begin{figure}[h]
\centering
\includegraphics[width=0.6\textwidth]{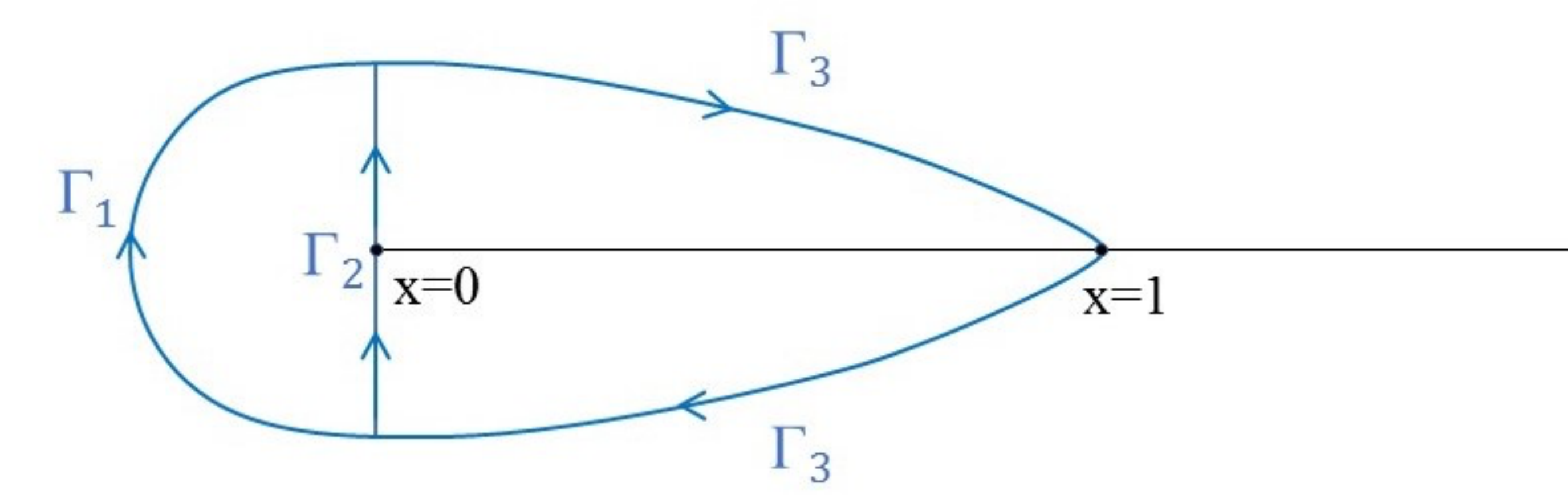}
\caption{The contours used in \reef{hsplit-cont}. It is important that $\Gamma_1$ goes around $x=0$, while $\Gamma_2$ passes through it.}
\label{Mazac-contour-mod}
\end{figure}

Additional complications arise near $x=0$. Namely, as a result of the infinite summation, the function $h(x)$ develops a cut over the negative real axis $x<0$. For this reason the contour in figure \ref{Mazac-contour} is no longer appropriate, and has to be modified.
In fact, the behavior of his $h(x)$ near $x=0$ can be described by the formula
\beq
h(x)=h_1(x)+h_2(x)\,,
\label{hsplit}
\eeq
where $h_1(x)$ is analytic near $x=0$, while $h_2(x)$ has a cut along $x<0$.
The total functional can then be defined as a sum of three terms
\beq
\omega(f)=\frac 1{2\pi i} \left(\int_{\Gamma_1} h_1(x) +\int_{\Gamma_2}h_2(x)+\int_{\Gamma_3}h(x)\right)\times(1-x)^{-2\Df} f(z(x))\,dx\,,
\label{hsplit-cont}
\eeq
where the three parts of the contour are chosen as in figure \ref{Mazac-contour-mod}. The finiteness and swapping conditions for 
this functional are \reef{L1-2} and \reef{L2-2} imposed on $h(x)$ and \reef{L2-1} imposed on $h_2(x)$, while $h_1(x)$ does not have to satisfy any condition near $x=0$. This contour prescription is equivalent to the one discussed in \cite{Mazac:2016qev} below Eq.~(5.19).

After this introduction, let's see how the functionals of \cite{Mazac:2016qev} fare with respect to all these conditions. To be more precise, his functionals correspond to
\beq
h(x)=\tilde h(x) +c(x)\,,
\eeq
where $\tilde h(x)$ is a sum as in \reef{seriesh} with summation over even/odd $n$ depending if $\Df$ is integer or halfinteger:
\begin{align}
\tilde h(x) &= \sum_{n\in 2\bN} a_n p_n(x)\qquad(\Df\in\bN)\,,\\
\tilde h(x) &= \sum_{n\in 2\bN-1} a_n p_n(x)\qquad(\Df\in\bN-\half)\,.
\end{align}
Using the properties of Legendre polynomials, this implies that
\begin{align}
\tilde h(1-x) &= -\tilde h(x)\qquad(\Df\in\bN)\,,\label{asym}\\
\tilde h(1-x) &= \tilde h(x)\qquad(\Df\in\bN-\half)\,.\label{sym}
\end{align}
On the other hand, $c(x)$ is a finite degree polynomial which can be used to make the total $h(x)$ vanish at $x=1$ sufficiently fast.

Consider first $\Df\in\bN$. In this case Maz\'a\v{c} says in section 5.2 (we translate his Eq.~(5.18) and others into our notation) that 
near $x=0$ we can decompose $\tilde h(x)$ as
\beq
\tilde h(x) = \tilde h_1(x)+\tilde h_2(x)\,,
\label{dectilde}
\eeq
where $\tilde h_1(x)$ is analytic near $x=0$, while $\tilde h_2(x)$ has a cut along $x<0$ and satisfies
\beq
\tilde h_2(x) =O(x^{2\Df} \log x)
\label{h2bound}
\eeq
Then by antisymmetry \reef{asym} we have the behavior near $x=1$:
\beq
\tilde h(x) = -\tilde h_1(1-x)-\tilde h_2(1-x)\,.
\eeq
In section 5.3 he uses the freedom to add $c(x)$ to set the behavior of 
\beq
-\tilde h_1(1-x)+c(x)=O((x-1)^{2\Df})\,.
\eeq
He conjectures that it's always possible although he only checked it up to $\Df=5$.
If so, we can define this functional as in \reef{hsplit}, \reef{hsplit-cont} with $h_1(x)=\tilde h_1(x)+c(x)$ and $h_2(x)=\tilde h_2(x)$ and have the conditions for the finiteness and swapping satisfied near both $x=0$ and $x=1$.

Consider next $\Df\in\bN-\half$, discussed in \cite{Mazac:2016qev}, section 5.4 and appendix A. For the particular values $\Df=1/2, 3/2, 5/2$ he provides explicit
$\tilde h(x)$, see his (4.29), (A.12)-(A.14). In these three cases we checked that near $x=0$ one has decomposition \reef{dectilde} with the nonanalytic part satisfying \reef{h2bound}. The behavior near $x=1$ is given by symmetry \reef{sym}:
\beq
\tilde h(x) = \tilde h_1(1-x)+\tilde h_2(1-x)\,.
\eeq
He says that he's able, at least for $\Df\le 9/2$, to use the freedom of adding $c(x)$ to set:
\beq
\tilde h_1(1-x)+c(x)=O((x-1)^{2\Df})\,.
\eeq
If that's the case then the conditions for the finiteness and swapping are indeed satisfied, just as for $\Df\in\bN$.

The bottom line is that in the cases of low-lying $\Df$, where Ref.~\cite{Mazac:2016qev} provides sufficient information, we are able to apply our criteria and to prove swapping. A more detailed understanding and an extension of his argument would be needed to establish this for general $\Df$. This is beyond the scope of our work.

\section{Conclusions}

Conformal field theories are both physically relevant and mathematically well defined. They satisfy precise axioms, which can be used to derive rigorous bounds separating the possible from the impossible. These bounds are usually argued by contradiction, employing the method of linear functionals. The non-explicit character of such arguments requires special care, otherwise one risks to throw out the baby with the bathwater. In this note we proposed a blueprint which needs to be followed to guarantee that this does not happen. As an application, we checked that the functionals recently constructed in \cite{Mazac:2016qev} can be safely used in the conformal bootstrap.

\section*{Acknowledgements}

JQ is grateful to the CERN Theoretical Physics Department for hospitality. SR is supported by Mitsubishi Heavy Industries as an ENS-MHI Chair holder, by the National Centre of Competence in Research SwissMAP funded by the Swiss National Science Foundation, and by the Simons Foundation grant 488655 (Simons collaboration on the Non-perturbative bootstrap).

\appendix
\section{Toy counterexample}
 \label{sec:toy}
 
Mathematics textbooks are full of examples when one cannot swap integration with summation. We give one here so that you don't have to go look for it yourself. The example is based on simple power series expansions. However, the mechanism is general, and one should beware of falling into similar traps when working with conformal block expansions. 

Consider the following functions on the real interval $0<t<1$:
\beq
\phi_0(t)= 1,\quad \phi_n(t)=(n+1) t^{n}- n t^{n-1}\quad(n=1,2\ldots)\,.
\eeq
The series of these functions sums to zero:
\beq
\sum_{n=0}^\infty \phi_n(t)=0\qquad(0<t<1)\,.
\label{toy}
\eeq
Indeed, it was designed so that the subsequent terms cancel telescopically, so that the partial sums
\beq
\sum_{n=0}^N \phi_n(t) = (N+1) t^{N} \to 0 \qquad(0<t<1)\,.
\eeq

Now consider formally integrating the series against some function $w(t)$:
\beq
\label{eq:wrong}
\sum_{n=0}^\infty I_n=0,\quad I_n= \int_0^1 dt\,w(t)\phi_n(t)\,.
\eeq
Let us check this in a couple of examples. If we take $w(t)=1-t$, then things work nicely:
\beq
I_0=\frac 12,\quad I_n=\frac{1}{n+2}-\frac{1}{n+1}\quad(n=1,2\ldots)\,,
\eeq
and the series in \reef{eq:wrong} does converge to zero. On the other hand, for $w(t)=1$ we have
\beq
I_0=1,\quad I_n =0\quad(n>0)\,,
\eeq
in manifest contradiction with \reef{eq:wrong}. 

To understand this ``paradox", consider the tails of the series \reef{toy}:
\beq
\Phi_N(t)=\sum_{n=N+1}^\infty \phi_n(t) = -(N+1) t^{N}\,.
\eeq
To swap integration and summation, we must have a condition analogous to \reef{Fsmall}:
\beq
\int dt\,w(t)\Phi_N(t) \to 0\qquad(N\to\infty)\,.
\eeq
This condition is satisfied for $w(t)=1-t$ but not for $w(t)=1$.

\section{Spectra with accumulation points}
\label{sec:cont} 

In the main text we made an assumption that the spectrum of operators appearing in the conformal block decomposition \reef{CBexp}
is discrete without accumulation points. However, there exist 2d and 1d CFTs with continuous spectrum, such as the Liouville theory and its associated boundary CFTs (although in $d>2$ there are no known examples showing such behavior). Here we will show that our main results remain unchanged if the spectrum is continuous or has accumulation points.

In such a general situation, Eq.~\reef{CBexp} should be replaced by an indefinite Stieltjes integral
\beq
\label{indefSt}
\calG(z)=\int_0^\infty dP(\Delta)\,G_\D(z)\,,
\eeq
associated with a monotonically increasing function $P(\Delta)$, $P(0)$=0. 
Convergence of this integral is understood in two steps. First one defines the integral for a finite upper limit:
\beq
\label{defSt}
\int_0^{\Delta_*} dP(\Delta)\,G_\D(z)\,.
\eeq
This is defined as the $N\to\infty$ limit of the Riemann-Stieltjes (RS) sums:
\beq
\sum_{i=0}^{N-1}[P(\Delta_{i+1})-P(\D_i)]\,G_{\D_i}(z)\,,
\eeq
corresponding to finer and finer subdivisions of the interval $[0,\Delta_*]$:
\beq
\Delta_0=0<\Delta_1<\ldots<\Delta_N=\Delta_*\,.
\eeq
Let $z$ vary over a region where $|\rho(z)|<1-\eps$. For such $z$, the functions $G_\Delta(z)$ depend uniformly continuously on $\Delta\in[0,\Delta_*]$. This is enough to guarantee that the RS sums have a uniform limit. Since the individual RS sums are analytic, their limit \reef{defSt} is analytic as well. 

The second step is to define the integral in \reef{indefSt} as the limit of \reef{defSt} as $\Delta_*\to\infty$. Since \reef{defSt} monotonically grows with $\Delta_*$ for $0<z<1$, the limit does exist on this interval. Then one argues as in section \ref{sec:anal}, using the property \reef{est} of conformal blocks, that the convergence as $\Delta_*\to\infty$ is uniform in the regions $|\rho(z)|<1-\eps$. This shows that the function $\calG(z)$ is analytic in the cut complex plane, just as before. 

By the given argument, we have the following approximation of $\calG(z)$ by finite sums of conformal blocks with two error terms:
\beq
\calG(z)= \sum_{i=0}^{N-1}[P(\Delta_{i+1})-P(\D_i)]\,G_{\D_i}(z)+\calG_{\Delta_*}(z)+\calG^{\rm RS}_{\Delta_*,N}(z)\,.
\eeq
The first error term $\calG_{\Delta_*}(z)$ is the difference between \reef{indefSt} and \reef{defSt}, while the second error term $\calG^{\rm RS}_{\Delta_*,N}(z)$ is the difference between \reef{defSt} and the RS sum. This is to be compared with the situation in the main text, where we had only the first error term.

The first error term has the same properties as before: it goes uniformly to zero with $\Delta_*\to\infty$ in the region $|\rho(z)|<1-\eps$, and it can be uniformly in $\Delta_*$ bounded by the full 4pt function, as in Eq.~\reef{estend}.

On the other hand, as discussed above, the second error term can be made uniformly small in the same region $|\rho(z)|<1-\eps$, by taking $N\to\infty$ (for any fixed $\Delta_*$). Outside of this region we can use a crude upper bound:
\beq
\label{crude1}
|\calG^{\rm RS}_{\Delta_*,N}(z)|\le const\left(1+ \log \frac{1}{1-|\rho(z)|}\right)\,,
\eeq
Here $const$ may depend on $\Delta_*$ but is independent of $N$. This bound follows from the fact that each individual conformal block satisfies such a bound.

Now we are in a position to repeat the analysis of section \ref{sec:func}. Eq.~\reef{interchange} is replaced by:
\beq
 \sum_{i=0}^{N-1}[P(\Delta_{i+1})-P(\D_i)]\,\omega(F_{\D_i})+\omega(\calF_{\Delta_*})+\omega(\calF^{\rm RS}_{\Delta_*,N})=0\,.
\eeq
When we take the limit $N\to\infty$ and then $\Delta_*\to\infty$, this will become the desired equation
\beq
\int_0^\infty dP(\Delta) \omega(F_\Delta)=0\,,
\eeq
provided that we can show \reef{Fsmall} (which is done exactly as before) and, in addition, that 
\beq
\label{eq:additional}
\omega(\calF^{\rm RS}_{\Delta_*,N})\to 0\quad (N\to\infty,\Delta_*\text{ fixed})\,.
\eeq
This extra condition is obvious for the simple functionals \reef{ex1},\reef{ex2} since $\calF^{\rm RS}_{\Delta_*,N}$ goes uniformly to zero in the relevant region of $z$. For the cut-touching functionals, a little thought has to be given to what happens near the points $x=0,1$.
Since this error term satisfies the same crude bound \reef{crude1} as the conformal blocks, one can recycle the estimates from section \ref{sec:fin}. Conditions \reef{L1-1}, \reef{L1-2} are then sufficient to guarantee \reef{eq:additional}.

The conclusion of this discussion is that the sufficient conditions for the finiteness and swapping derived in the main text remain valid when the spectrum is continuous or discrete with accumulation points.
\small

\providecommand{\href}[2]{#2}\begingroup\raggedright\endgroup



\begin{thebibliography}{10}

\bibitem{Ferrara:1973yt}
S.~Ferrara, A.~F. Grillo, and R.~Gatto, ``{Tensor representations of conformal
  algebra and conformally covariant operator product expansion},''
\href{http://dx.doi.org/10.1016/0003-4916(73)90446-6}{{\em Annals Phys.}
  {\bfseries 76} (1973) 161--188}.

\bibitem{Polyakov:1974gs}
A.~M. Polyakov, ``{Nonhamiltonian approach to conformal quantum field
  theory},''
{\em Zh. Eksp. Teor. Fiz.} {\bfseries 66} (1974) 23--42.

\bibitem{Mack:1975jr}
G.~Mack, ``{Duality in quantum field theory},''
\href{http://dx.doi.org/10.1016/0550-3213(77)90238-3}{{\em Nucl. Phys.}
  {\bfseries B118} (1977) 445--457}.

\bibitem{Belavin:1984vu}
A.~A. Belavin, A.~M. Polyakov, and A.~B. Zamolodchikov, ``{Infinite conformal
  symmetry in two-dimensional quantum field theory},''
\href{http://dx.doi.org/10.1016/0550-3213(84)90052-X}{{\em Nucl. Phys.}
  {\bfseries B241} (1984) 333--380}.

\bibitem{Rattazzi:2008pe}
R.~Rattazzi, V.~S. Rychkov, E.~Tonni, and A.~Vichi, ``{Bounding scalar operator
  dimensions in 4D CFT},''
  \href{http://dx.doi.org/10.1088/1126-6708/2008/12/031}{{\em JHEP} {\bfseries
  12} (2008) 031},
\href{http://arxiv.org/abs/0807.0004}{{\ttfamily arXiv:0807.0004 [hep-th]}}.

\bibitem{Rychkov:2016iqz}
S.~Rychkov, \href{http://dx.doi.org/10.1007/978-3-319-43626-5}{{\em {EPFL
  Lectures on Conformal Field Theory in $D\ge 3$ Dimensions}}}.
\newblock SpringerBriefs in Physics. 2016.
\newblock
\href{http://arxiv.org/abs/1601.05000}{{\ttfamily arXiv:1601.05000 [hep-th]}}.
\newblock

\bibitem{Simmons-Duffin:2016gjk}
D.~Simmons-Duffin, \href{http://dx.doi.org/10.1142/9789813149441_0001}{``{TASI
  Lectures on the Conformal Bootstrap},''} in {\em {Proceedings, TASI: Boulder,
  CO, USA, June 1-26, 2015}}, pp.~1--74.
\newblock 2017.
\newblock \href{http://arxiv.org/abs/1602.07982}{{\ttfamily arXiv:1602.07982
  [hep-th]}}.
\newblock
\url{https://sites.google.com/a/colorado.edu/tasi-2015-wiki/lecture-topics/conformal-bootstrap}.
\newblock

\bibitem{DO1}
F.~Dolan and H.~Osborn, ``{Conformal four point functions and the operator
  product expansion},''
  \href{http://dx.doi.org/10.1016/S0550-3213(01)00013-X}{{\em Nucl.Phys.}
  {\bfseries B599} (2001) 459--496},
\href{http://arxiv.org/abs/hep-th/0011040}{{\ttfamily arXiv:hep-th/0011040
  [hep-th]}}.

\bibitem{DO2}
F.~Dolan and H.~Osborn, ``{Conformal partial waves and the operator product
  expansion},'' \href{http://dx.doi.org/10.1016/j.nuclphysb.2003.11.016}{{\em
  Nucl.Phys.} {\bfseries B678} (2004) 491--507},
\href{http://arxiv.org/abs/hep-th/0309180}{{\ttfamily arXiv:hep-th/0309180
  [hep-th]}}.

\bibitem{DO3}
F.~Dolan and H.~Osborn, ``{Conformal Partial Waves: Further Mathematical
  Results},''
\href{http://arxiv.org/abs/1108.6194v2}{{\ttfamily arXiv:1108.6194v2
  [hep-th]}}.

\bibitem{Echeverri:2016ztu}
A.~Castedo~Echeverri, B.~von Harling, and M.~Serone, ``{The Effective
  Bootstrap},'' \href{http://dx.doi.org/10.1007/JHEP09(2016)097}{{\em JHEP}
  {\bfseries 09} (2016) 097},
\href{http://arxiv.org/abs/1606.02771}{{\ttfamily arXiv:1606.02771 [hep-th]}}.

\bibitem{Mazac:2016qev}
D.~Maz\'a\v{c}, ``{Analytic Bounds and Emergence of $\textrm{AdS}_2$ Physics
  from the Conformal Bootstrap},''
  \href{http://dx.doi.org/doi:10.1007/JHEP04(2017)146}{{\em JHEP} {\bfseries
  04} (2017) 146},
\href{http://arxiv.org/abs/1611.10060}{{\ttfamily arXiv:1611.10060 [hep-th]}}.

\bibitem{Mazac-seminar}
D.~Maz\'a\v{c}.
\newblock \url{https://www.youtube.com/watch?v=Q1Tefm1FQhA}. Simons Bootstrap
  Collaboration Online Seminar (Jan 18, 2016).

\bibitem{Pappadopulo:2012jk}
D.~Pappadopulo, S.~Rychkov, J.~Espin, and R.~Rattazzi, ``{OPE Convergence in
  Conformal Field Theory},''
  \href{http://dx.doi.org/10.1103/PhysRevD.86.105043}{{\em Phys.Rev.}
  {\bfseries D86} (2012) 105043},
\href{http://arxiv.org/abs/1208.6449}{{\ttfamily arXiv:1208.6449 [hep-th]}}.

\bibitem{Hogervorst:2013sma}
M.~Hogervorst and S.~Rychkov, ``{Radial Coordinates for Conformal Blocks},''
  \href{http://dx.doi.org/10.1103/PhysRevD.87.106004}{{\em Phys.Rev.}
  {\bfseries D87} (2013) 106004},
\href{http://arxiv.org/abs/1303.1111}{{\ttfamily arXiv:1303.1111 [hep-th]}}.

\end{thebibliography}
\end{document}